\documentclass[aps,prc,twocolumn,showpacs,superscriptaddress]{revtex4-1}
\usepackage[english]{babel}
\usepackage[T1]{fontenc}
\usepackage{graphicx}
\usepackage{subfigure}
\usepackage{bm,amssymb,amsmath}
\usepackage[colorlinks=true,linkcolor=blue,citecolor=blue,urlcolor=blue]{hyperref}
\usepackage{booktabs}
\usepackage{multirow}
\usepackage{color}

\newcommand{\goodgap}{%
	\hspace{\subfigtopskip}%
	\hspace{\subfigbottomskip}}

\begin{document}

\title{Effects of the two-body and three-body hyperon-nucleon interactions in $\Lambda$ hypernuclei}

\author{D. Lonardoni}
\affiliation{Dipartimento di Fisica, Universit\`a di Trento, via Sommarive, 14 I-38123 Trento, Italy}
\affiliation{INFN - Istituto Nazionale di Fisica Nucleare, Gruppo Collegato di Trento, I-38123 Trento, Italy}

\author{S. Gandolfi}
\affiliation{Theoretical Division, Los Alamos National Laboratory, Los Alamos, New Mexico 87545, USA}

\author{F. Pederiva}
\affiliation{Dipartimento di Fisica, Universit\`a di Trento, via Sommarive, 14 I-38123 Trento, Italy}
\affiliation{INFN - Istituto Nazionale di Fisica Nucleare, Gruppo Collegato di Trento, I-38123 Trento, Italy}

\begin{abstract}
\noindent{\bf Background:} The calculation of the hyperon binding energy in hypernuclei is
crucial to understanding the interaction between hyperons and nucleons.

\noindent{\bf Purpose:} We assess the relative importance of two- and three-body hyperon-nucleon
force by studying the effect of the hyperon-nucleon-nucleon interaction
in closed shell $\Lambda$ hypernuclei from $A=5$ to $91$.

\noindent{\bf Methods:} The $\Lambda$ binding energy has been calculated using the
auxiliary field diffusion Monte Carlo method for the first time, to
study light and heavy hypernuclei within the same model.

\noindent{\bf Results:} Our results show that including a three-body component in
the hyperon-nucleon interaction leads to a saturation of the $\Lambda$
binding energy remarkably close to the experimental data. In contrast,
the two-body force alone gives an unphysical limit for the binding energy.

\noindent{\bf Conclusions:} The repulsive contribution of the three-body hyperon-nucleon-nucleon
force is essential to reproduce, even qualitatively, the binding energy of the
hypernuclei in the mass range considered.
\end{abstract}

\pacs{21.80.+a, 26.60.Kp, 21.60.De}

\maketitle

The onset of strange baryons ($\Sigma^{-}$ and $\Lambda$) in neutron
matter at densities of order $(2-3)\rho_0$, where $\rho_0=0.16$~fm$^{-3}$,
has been questioned for a long time. 
Recent theoretical calculations based on the Brueckner-Hartree-Fock 
theory suggest that
any process generating new on-shell degrees of freedom in
high-density fermionic matter leads to a substantial softening of its
equation of state (EOS) (see, for example, \cite{Vidana:2011,Schulze:2011}
and references therein). When occurring in the inner core of a neutron
star, such a mechanism would reduce the value of its predicted maximum
mass and of its radius. Until a few years ago, astrophysical observations of
neutron stars were concentrated in a relatively narrow region in the
neighborhood of the Chandrasekhar limit ($M\simeq~1.41M_\odot$). Most of
the realistic EOSs based on the hypothesis that matter is made of nucleons
only, while compatible with these observations, predict a maximum mass
typically larger than 2$M_\odot$. This result can be considered very
robust. As an example, recent quantum Monte Carlo (QMC) calculations of
the equation of state of pure neutron matter (PNM), symmetric nuclear
matter (SNM) and baryonic matter at $\beta$ and $\mu$ equilibrium using
realistic density-dependent potentials (DDPs)~\cite{Gandolfi:2010},
essentially confirm the behavior predicted by Akmal, Pandharipande 
and Ravenhall with a full AV18+three-body interaction~\cite{Akmal:1998}. 
With such a nuclear Hamiltonian the predicted EOS supports a maximum neutron 
star mass larger than 1.97$M_\odot$ recently observed~\cite{Demorest:2010}.

In so far as the appearance of strange baryons is concerned,
the situation is more controversial. Some authors (see,
e.g.,~\cite{Bednarek:2012,Weissenborn:2012}), suggest that
the appearance of hyperons in the EOS does
not lead to very strong effects. Other recent papers, like
Refs.~\cite{Dapo:2010,Schulze:2011,Massot:2012,Vidana:2011,Miyatsu:2012},
show a more substantial influence, but with contradictory outcomes in
terms of the predicted maximum mass of neutron stars not compatible
with the observations~\cite{Steiner:2012}. Therefore, the issue is far
from being completely settled.

A combination of reasons leads to the uncertainty in the analysis of
the influence of strangeness degrees of freedom in the EOS. First of all,
the interaction between nucleons and hyperons is still far from being
known with sufficient accuracy. The prospective measurements of properties of
light hypernuclei should
improve the quality of the available data, making possible a realistic
phenomenological analysis. Second, the theoretical tools employed are
all affected by uncontrollable intrinsic approximations as soon as one
tries to push the study beyond few-body systems. As a consequence,
so far it is not clear how well the model hyperon-nucleon ($YN$)
potentials work in the limit of medium mass hypernuclei, and,
as a consequence, in the extrapolation to homogeneous matter. 
However, in the last few years important advances have been made both
on the experimental and on the theoretical side. Several experiments
aim to measure the binding energy of different 
$\Lambda$ hypernuclei~\cite{Saito:2012,Sato:2011,Garibaldi:2013}. 
On the theoretical side, the development
of quantum Monte Carlo methods has opened the way to study consistently
nuclear systems from few nucleons to infinite matter~\cite{Pieper:2001,
Gandolfi:2009,Gandolfi:2011} within the same scheme or model.

In this Rapid Communication we discuss the use the auxiliary field diffusion Monte Carlo (AFDMC) model , 
to study a nonrelativistic Hamiltonian
based on a phenomenological $\Lambda N$ interaction in order to show how the
inclusion of explicit $\Lambda NN$ terms provides the necessary repulsion to
realistically describe the separation energy of a $\Lambda$ hyperon
in hypernuclei of intermediate masses. This point makes very clear the
fact that the lack of an accurate Hamiltonian might be responsible for
the unrealistic predictions of the EOSs that would tend to rule out the
appearance of strange baryons in high-density matter.

\begin{figure*}[!t]
	\begin{center}
		\subfigure[\label{fig:LN_2pi}]{\includegraphics[height=3.2cm]{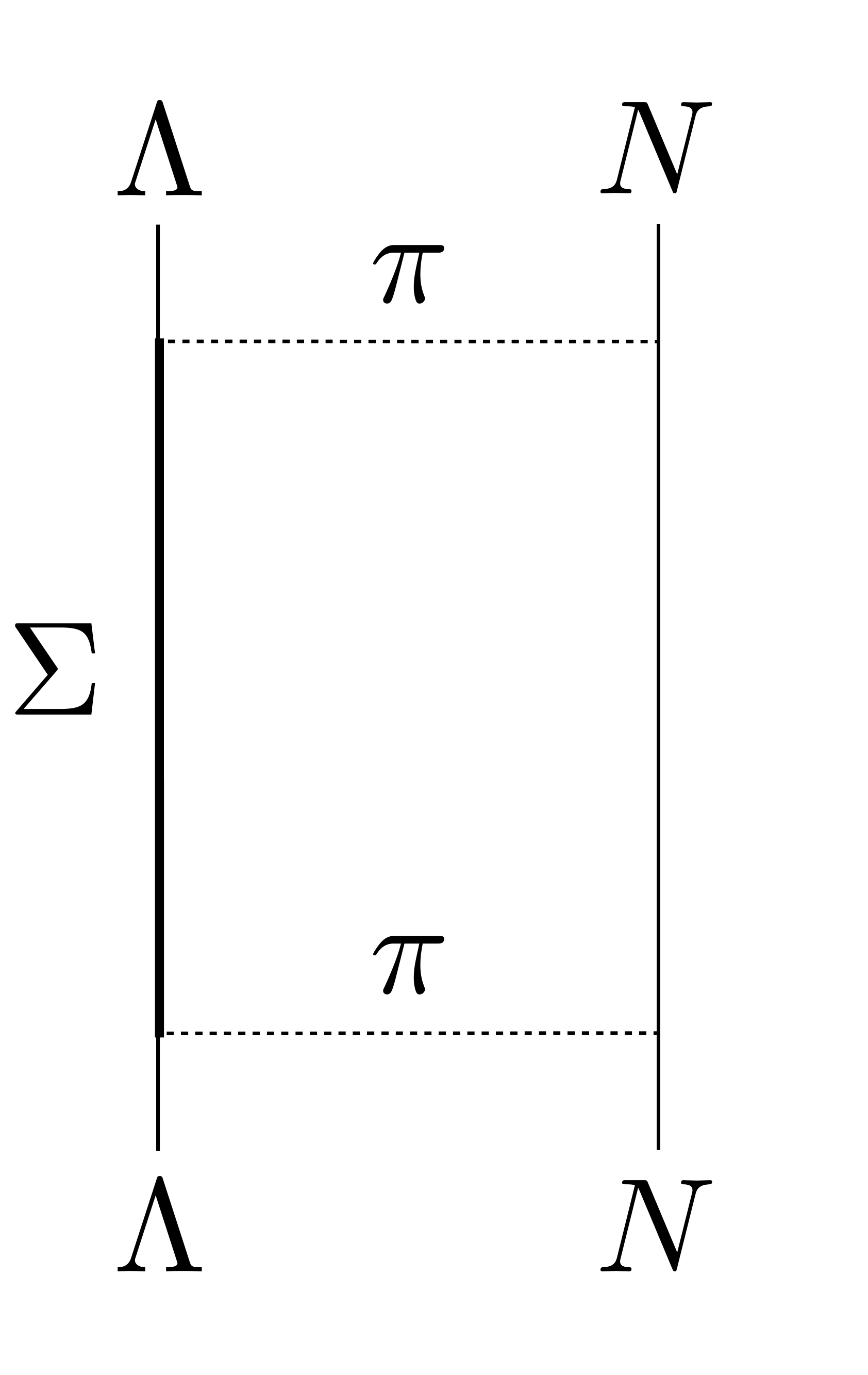}}
		\goodgap\goodgap\goodgap
		\subfigure[\label{fig:LN_K}]{\includegraphics[height=3.2cm]{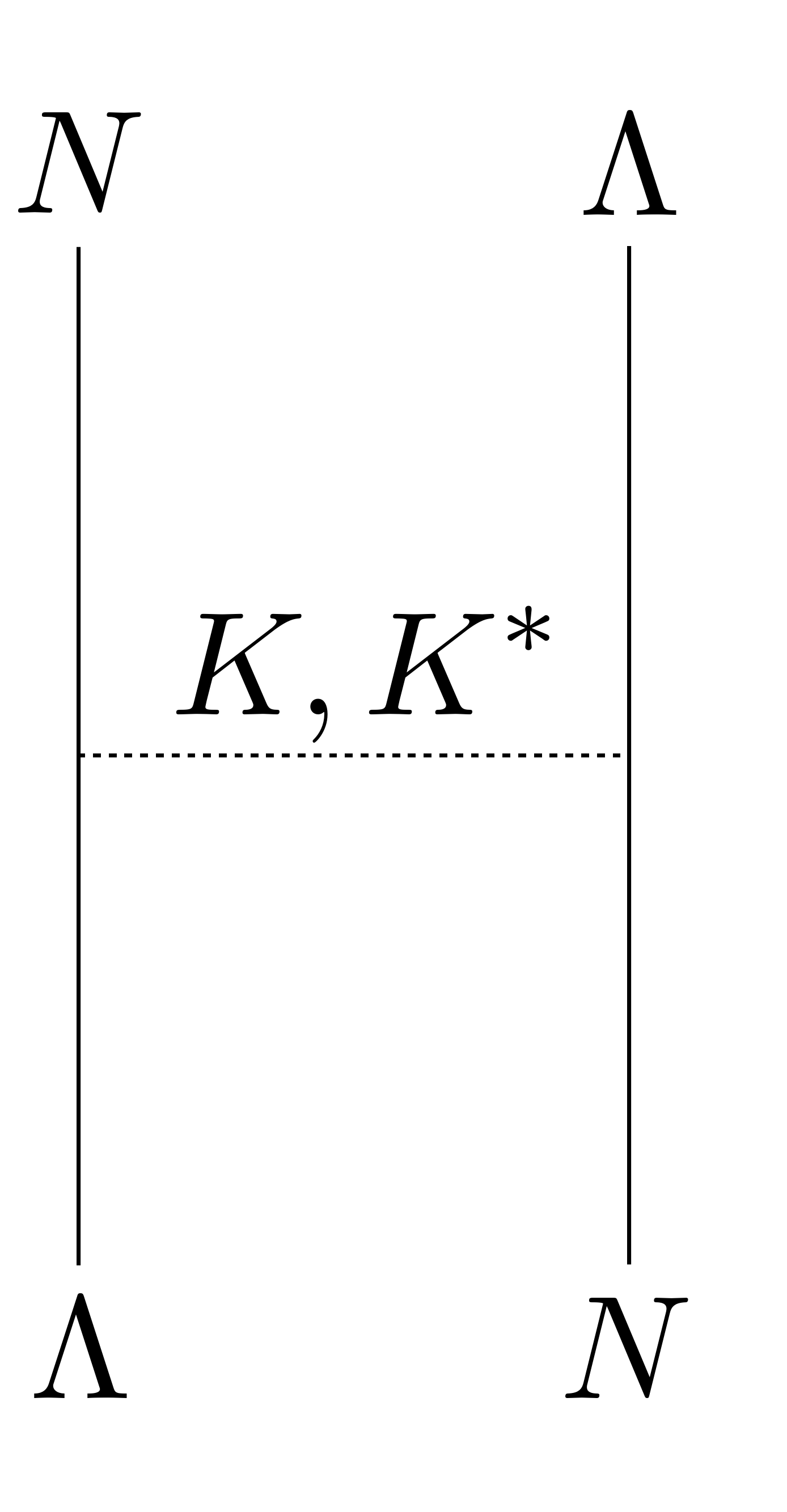}}
		\goodgap\goodgap\goodgap
		\subfigure[\label{fig:LNN_pw}]{\includegraphics[height=3.2cm]{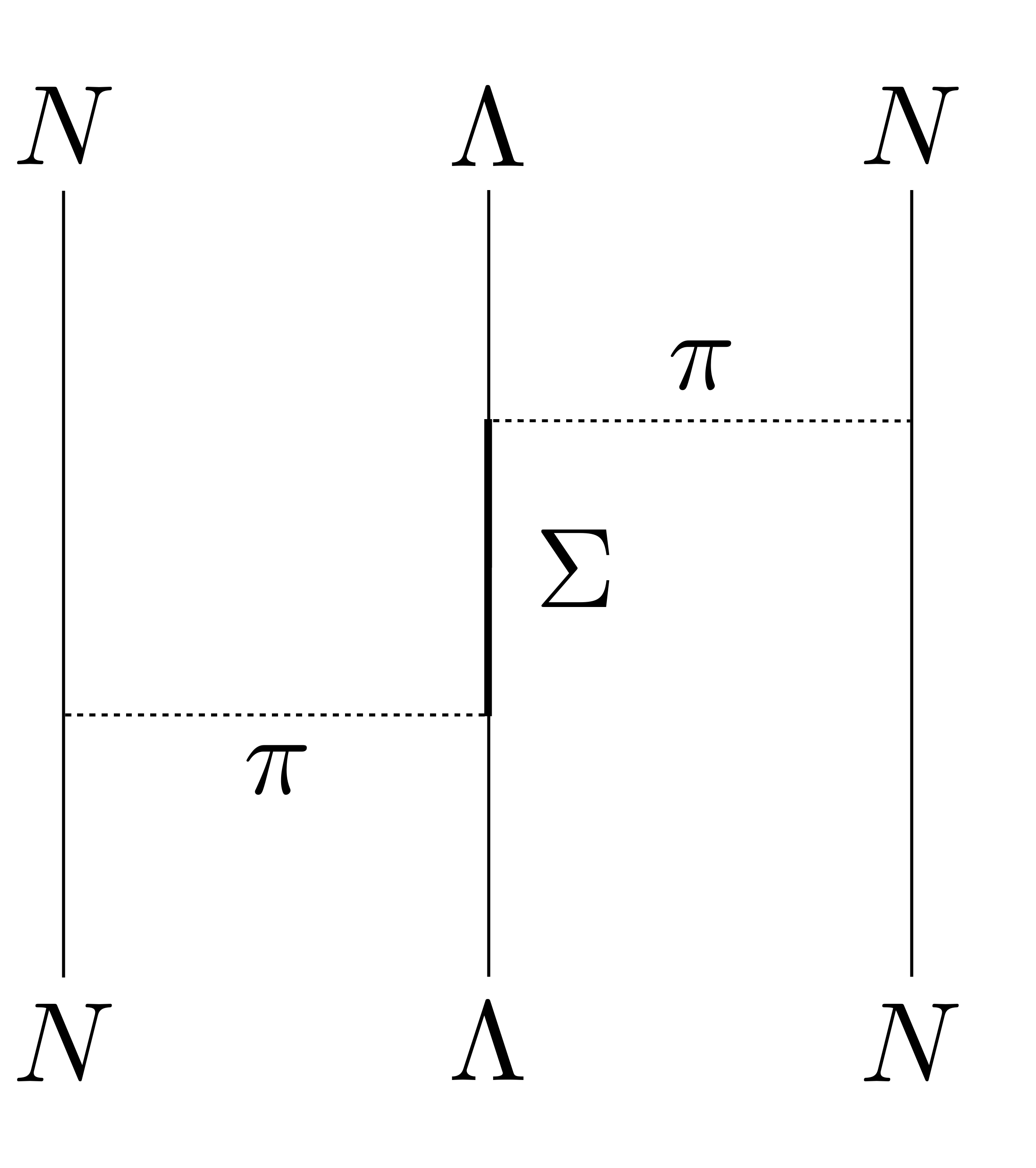}}
		\goodgap\goodgap\goodgap
		\subfigure[\label{fig:LNN_sw}]{\includegraphics[height=3.2cm]{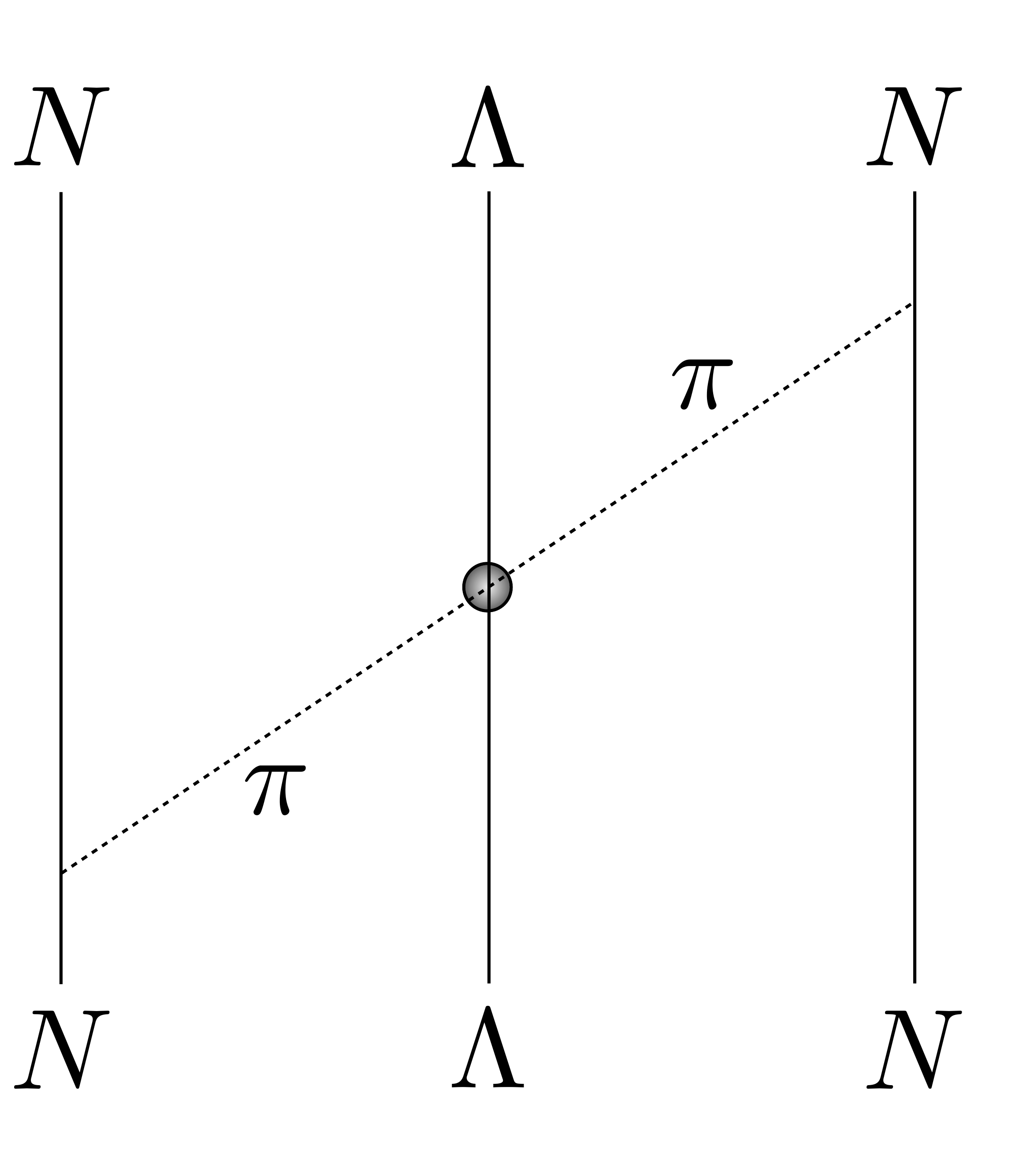}}
		\goodgap\goodgap\goodgap
		\subfigure[\label{fig:LNN_d}]{\includegraphics[height=3.2cm]{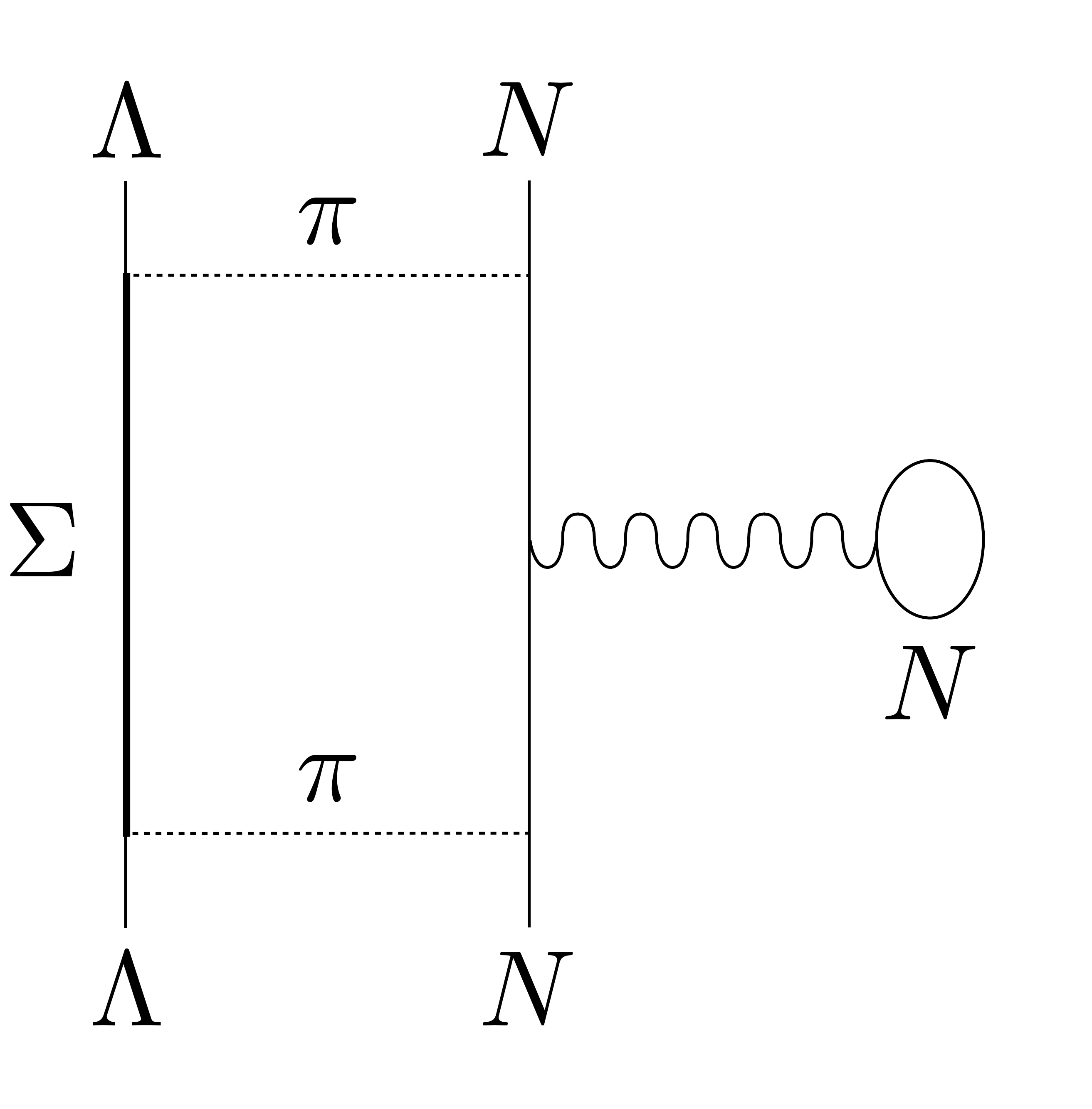}}
	\caption{Meson exchange processes between nucleons and hyperons. 
		\ref{fig:LN_2pi} and \ref{fig:LN_K} represent the $\Lambda N$ channels.
		\ref{fig:LNN_pw}-\ref{fig:LNN_d}
		are the three-body $\Lambda NN$ channels included in the potential by Usmani 
		\emph{et al.}~\cite{Usmani:1995,Usmani:2008}.}
	\label{fig:diagrams} 
	\end{center}
\end{figure*}

After the pioneering work reported in Ref.~\cite{Dalitz:1972},
several models have been proposed to describe the hyperon-nucleon
($YN$) interaction. A number of potentials in the Nijmegen soft-core
form have been developed in the past (like NSC89 and NSC97x). A recent
review of these interactions, together with Hartree-Fock (HF) calculations have been
published by Djapo \emph{et al.}~\cite{Dapo:2008}. These potentials are
accurate in describing the available scattering two-body data,
and have been used in the BHF studies previously quoted. Starting
in the 1980s, a class of Argonne-like interactions have been
developed by Bodmer, Usmani, and Carlson on the grounds of quantum
Monte Carlo calculations. A possible form of
a three-body $YNN$ interaction was also proposed in the same context
~\cite{Bodmer:1984,Bodmer:1988,Usmani:1995,Usmani:1999,Usmani:2008}. More
recently Polinder \emph{et al.}~\cite{Polinder:2006} proposed a potential based
on a chiral perturbation theory expansion. As an alternative a cluster
model to study light hypernuclei has been recently proposed by 
Hiyama and collaborators (see, for example, Refs.~\cite{Hiyama:2010,Hiyama:2009}).
Interesting results on $\Lambda$ hypernuclei have also been obtained 
within a $\Lambda$-nucleus potential model, in which the need of a functional 
with a more than linear density dependence was shown, suggesting the 
importance of a many-body interaction ~\cite{Millener:1988}.
Finally, other methods based on mean-field techniques have been used
to study medium and heavy mass hypernuclei~\cite{Vidana:1998,Keil:2000}.

An important point that needs to be made is that $\Lambda N$ and $\Lambda NN$
interactions are both at the two-pion exchange (TPE) order. Another important 
difference with respect the nucleonic case is that the mass of the intermediate 
excited state $\Sigma$ compared to the $\Lambda$ is much smaller than
in the pure nucleonic case, where the difference between the nucleon and 
the $\Delta$ is much larger.
$\Lambda N$ and $\Lambda NN$
interactions should therefore be considered necessary in any consistent
theoretical calculation. In 2002 Nogga \emph{et al.}~\cite{Nogga:2002} performed
Fadeev-Yakubowsky calculations of the $0^{+}$ and the $1^{+}$ state
of $^4_\Lambda$H and $^4_\Lambda$He in order to study charge symmetry
breaking effects. In both cases they predict a $\Lambda$-separation
energy that is too small and claim that an attractive $\Lambda NN$ interaction is
necessary. 

We have revisited the problem from a slightly different starting point. We
employed a potential in coordinates space, including an explicit 
repulsion between baryons, with $NN$, $\Lambda N$ and $\Lambda NN$ components. Keeping the
parameters of the $\Lambda N$ interaction at the values determined by Usmani
\emph{et al.}, we computed the ground-state energy of a set of hypernuclei,
and calculate for each the quantity $B_\Lambda$, i.e., the separation
energy of the $\Lambda$ hyperon, by means of the 
AFDMC method, using a realistic nucleon-nucleon
interaction.  We select one of the possible set of parameters of the
$\Lambda NN$ interaction suggested in Bodmer \emph{et al.}, 
and then by Usmani and collaborators, that reasonably
reproduces experimental data on a set of light hypernuclei.

Within this model, nuclei and hypernuclei are described as
nonrelativistic particles interacting via two-
and three-body forces:
\begin{align}
	H_{\text{nuc}}&=\sum_{i=1}^{A-1}\frac{p_i^2}{2m_N}+\sum_{i<j}^{A-1}v_{ij} \,,\\
	H_{\text{hyp}}&=H_{\text{nuc}}+\frac{p_\Lambda^2}{2m_\Lambda}+\sum_{i=1}^{A-1}v_{\Lambda i}+\sum_{i<j}^{A-1}v_{\Lambda ij}\,.
\end{align}
Here $A$ refers to the total number of baryons, nucleons plus the
$\Lambda$ particle. To test the effect of using different
nuclear Hamiltonians on the $\Lambda$-separation energy, and to test
the compatibility of the $NN$ interaction with the $\Lambda N$ and $\Lambda NN$
forces, we use
three different two-nucleon potentials $v_{ij}$: the Argonne AV4'
and AV6'~\cite{Wiringa:2002}, that are simplified versions of the
Argonne AV18~\cite{Wiringa:1995} and the Minnesota potential 
from Ref.~\cite{Thompson:1977}.

Isospin conservation implies that a $\Lambda$ hyperon can exchange
a pion only with a $\Lambda\pi\Sigma$ vertex. This fact leads to the
consequence that one-pion exchange (OPE) processes are not allowed. The
lowest order $\Lambda N$ coupling must therefore involve the exchange
of two pions, with the formation of a virtual $\Sigma$ hyperon, as
illustrated in Fig.~\ref{fig:LN_2pi}. One-meson exchange processes can
only occur through the exchange of a $K$ or $K^*$. This process
has the effect of exchanging the strangeness between the two baryons,
as shown in Fig.~\ref{fig:LN_K}. The $\Lambda N$ interaction can therefore be
modeled with a central term, which includes the $\Lambda N$ exchange
operator $\varepsilon(\hat P_x-1)$, plus a spin-dependent contribution:
\begin{equation}
	v_{\Lambda i}=v_{0}(r_{\Lambda i})(1-\varepsilon+\varepsilon \hat{P}_x)+\tfrac{1}{4}v_\sigma T^2_\pi(m_\pi r_{\Lambda i})\,{\bm\sigma}_\Lambda\cdot{\bm\sigma}_i\,,
	\label{eq:V_YN}
\end{equation}
where $\hat P_x$, $v_0$ and $T_\pi^2$ are defined in Ref.~\cite{Usmani:2008} 
and references therein, and ${\bm\sigma}_\Lambda$ and ${\bm\sigma}_i$ are Pauli 
matrices acting on the $\Lambda$ and nucleons.
Both the spin-dependent and the central terms contain the usual tensor
operator $T_\pi$ acting twice. All the pion exchange interaction is
therefore active at intermediate range. The short-range contributions
are as usual included by means of a phenomenological central repulsive
factor, included in $v_0(r)$. For more details see, for example,
Ref.~\cite{Usmani:2008}.

The remaining diagrams in Fig.~\ref{fig:diagrams} are two-nucleon
one-hyperon interactions, which are at the same TPE order, and should
therefore be included together with the two-body part in order to
have a consistent description. The three-body potential $v_{\Lambda ij}$
can be conveniently decomposed in a contribution that we label
as $v^{2\pi}_{\Lambda ij}=v^{P}_{\Lambda ij}+v^{S}_{\Lambda ij}$,
and that corresponds to the $p$-wave and $s$-wave two-pion exchange
diagrams (respectively \ref{fig:LNN_pw} and \ref{fig:LNN_sw}), 
and a dispersive term that includes short-range
contributions, labeled as $v_{\Lambda ij}^{D}$. They can be expressed as:
\begin{align}
	v_{\Lambda ij}^{D}&=W^{D}T_{\pi}^{2}\left(m_{\pi}r_{\Lambda i}\right)T^{2}_{\pi}\left(m_{\pi}r_{\Lambda j}\right)
		\!\!\left[1+\frac{1}{6}{\bm\sigma}_\Lambda\!\cdot\!\left({\bm\sigma}_{i}+{\bm\sigma}_{j}\right)\right]\,,\nonumber \\
	v_{\Lambda ij}^{P}&=-\left(\frac{C^{P}}{6}\right)\left({\bm\tau}_{i}\cdot{\bm\tau}_{j}\right)
	\Bigl\{X_{i\Lambda}\,,X_{\Lambda j}\Bigr\}\label{eq:V_YNN} \,,\\[0.5em]
	v_{\Lambda ij}^{S}&=C^{S}Z\left(m_{\pi}r_{\Lambda i}\right)Z\left(m_{\pi}r_{\Lambda j}\right)
		\!\left({\bm\sigma}_{i}\cdot\hat{\bm r}_{i\Lambda}\;{\bm\sigma}_{j}\cdot\hat{\bm r}_{j\Lambda}\right){\bm\tau}_{i}\cdot{\bm\tau}_{j}\nonumber \,.
\end{align}
The definition of the functions $X_{i\Lambda}$ and $Z(x)$ as well
as the range of parameters for the three-body force can be found in
\cite{Usmani:2008} and references therein.

The ground-state energy of the many-body nuclear and hypernuclear
Hamiltonians is computed by means of the AFDMC method. The algorithm
was originally introduced by Schmidt and Fantoni~\cite{Schmidt:1999}
in order to deal in an efficient way with spin-dependent Hamiltonians.
A trial wave function $\Psi_T$ is propagated in imaginary-time $\tau$
by sampling configurations of the system in coordinate-spin-isospin
space. Expectation values are computed averaging over the sampled
configurations. In the $\tau\rightarrow\infty$ limit, the evolved state
approaches the ground-state of $H$ and thus the ground-state properties
of the system can be obtained.

For a system with $A$ nucleons, the quadratic operator structure
$O_n^2$ of the nuclear Hamiltonians leads to a number of
spin-isospin states in the propagated wave function which grows 
exponentially with $A$.
This number quickly becomes intractable as $A$ gets large. Standard
Green's function Monte Carlo (GFMC) calculations are in fact limited to up to
12 nucleons~\cite{Pieper:2005} or 16 neutrons~\cite{Gandolfi:2011}.  
By applying the Hubbard-Stratonovich
transformation the computational cost of
the calculation becomes proportional to $A^3$ and systems with a larger
number of particles can be studied~\cite{Gandolfi:2009}. 
The AFDMC algorithm can be applied
to nuclear systems interacting via the Argonne V6-type potentials,
for which the two-body force can be separated into a spin-independent
and a spin-dependent part. The latter can be written as a sum of real
matrices which contain proper combinations of the components of V6. By
means of the diagonalization of such matrices it is possible to write the
imaginary-time propagator in the Hubbard-Stratonovich form
(see Refs.~\citep{Gandolfi:2006,Gandolfi:2007,Gandolfi:2009} for a
detailed discussion).
However, a realistic three-body force cannot be included in the propagator.

A straightforward variant of AFDMC can be applied to
$\Lambda$-hypernuclear systems, including the two-body~[Eq.~\ref{eq:V_YN}]
and three-body~[Eq.~\ref{eq:V_YNN}] hyperon-nucleon interactions.
It is indeed possible to recast the $\Lambda N$
and $\Lambda NN$ interactions so that they contain at most two-body
operators. These terms can directly be included in the AFDMC propagator.
The rest of the algorithm closely follows the
nucleon-only version~\cite{Gandolfi:2009}.

We assume that the wave function of a single $\Lambda$ hypernucleus is
a nuclear Slater determinant (the same as in Ref.~\citep{Gandolfi:2007}),
multiplied by a single particle wave function for the $\Lambda$
hyperon. For nucleon single-particle states we use the radial solutions of
the Hartree-Fock problem with the Skyrme force and we consider a $1s_{1/2}$
single-particle state for the $\Lambda$ particle. With the wave
function defined we consider nucleons and the hyperon as distinct particles. In
this way, we do not include the $\Lambda N$ exchange term of the $\Lambda N$
potential directly in the AFDMC propagator, because it mixes hyperon
and nucleon states. A perturbative treatment of this factor is, however,
possible.

A direct comparison of energy calculations with experimental results is 
given for the $\Lambda$-separation energy, defined as:
\begin{equation}
	B_\Lambda = B_{\text{nuc}}-B_{\text{hyp}},
\end{equation}
where $B_{\text{nuc}}$ and $B_{\text{hyp}}$ are, respectively, the total binding energies
of a nucleus with $A$ nucleons and the corresponding hypernucleus
with $A$ nucleons plus one $\Lambda$. The most significant outcome
of the calculation is the fact that the inclusion of the three-body
$\Lambda NN$ interaction qualitatively changes the saturation properties of
the $\Lambda$-separation energy. However, this result might depend
on the particular choice of the $NN$ interaction used to describe both
the nucleus and the hypernucleus. In particular, one might expect a
strong influence from the different nucleon density generated by
disparate models. In order to discuss this possible dependence, we
performed calculations with different $NN$ interactions having very
different saturation properties. The nuclear Hamiltonians considered 
here are semirealistic and can be easily implemented within the 
AFDMC scheme. We should point out that in neither case did we use a 
three-nucleon interaction.

In Tab.~\ref{tab:He5L-O17L} we show the results of the AFDMC
simulations for the $\Lambda$-separation energy in $^5_\Lambda$He and
$^{17}_{~\Lambda}$O.
For each hypernucleus, the two columns correspond to calculations using
the $\Lambda N$ interaction only or both the $\Lambda N$+$\Lambda NN$ force
of Ref.~\cite{Usmani:2008} with different $NN$ interactions. 
As it can be seen, for $^5_\Lambda$He
the extrapolated values of $B_\Lambda$ with the two-body $\Lambda N$ interaction
alone are about 10\% off and well outside statistical errors. In contrast 
the inclusion of the three-body $\Lambda NN$ force gives a similar $\Lambda$
binding energy independently to the choice of the $NN$ force.
On the grounds of this observation, we feel confident that the use of AV4', that makes
AFDMC calculations less expensive and more stable, will in any case
return realistic estimates of $B_\Lambda$ for larger masses
when including the $\Lambda NN$ interaction. We checked this assumption
performing simulations in $^{17}_{~\Lambda}$O, where the discrepancy
between the $\Lambda$-separation energy computed using the different
$NN$ interactions and the full $\Lambda N$+$\Lambda NN$ force is less than few per cent
(last column of Tab.~\ref{tab:He5L-O17L}).
The various $NN$ forces considered here are quite different. The AV6'
includes a tensor force, while AV4' and Minnesota have a simpler 
structure. We compared the AV4' and Minnesota, which have a similar 
operator structure but very different intermediate- and short-range 
correlations. 
The fact that the inclusion of the $\Lambda NN$ force does not depend
too much on the nuclear Hamiltonian is quite remarkable, because the
different $NN$ forces produce a quite different saturation point for the
nuclear matter EOS, suggesting that our results are pretty robust.
The discrepancies between our results and the experimental data are
likely due to the $\Lambda NN$ force that could be improved, while
the term due to $K$ exchange not included in our calculation are expected
to be small.

\renewcommand{\arraystretch}{1.2}
\begin{table}[!ht]
	\begin{center}
		\begin{tabular*}{\linewidth}{@{\hspace{0.5em}\extracolsep{\fill}}lcccc@{\extracolsep{\fill}\hspace{0.5em}}}
			\toprule
			\toprule
			\multirow{2}{*}{$NN$ potential} & \multicolumn{2}{c}{$^5_\Lambda$He}  & \multicolumn{2}{c}{$^{17}_{~\Lambda}$O} \\
			\cmidrule(l){2-3}\cmidrule(l){4-5}
			& \hspace{1em}$V_{\Lambda N}$ & $V_{\Lambda N}$+$V_{\Lambda NN}$ & \hspace{1em}$V_{\Lambda N}$ & $V_{\Lambda N}$+$V_{\Lambda NN}$ \\
			\midrule
			Argonne V4'\ &     \hspace{1em} 7.1(1)     &             5.1(1)               &   \hspace{1em} 43(1)        &        19(1) \\
			Argonne V6'\ &     \hspace{1em} 6.3(1)     &             5.2(1)               &   \hspace{1em} 34(1)        &        21(1) \\
			Minnesota\   &     \hspace{1em} 7.4(1)     &             5.2(1)               &   \hspace{1em} 50(1)        &        17(2) \\
			Expt.          & \multicolumn{2}{c}{3.12(2)} & \multicolumn{2}{c}{13.0(4)} \\
			\bottomrule
			\bottomrule
		\end{tabular*}
	\caption{$\Lambda$-separation energies (in MeV) for
	$^5_\Lambda$He and $^{17}_{~\Lambda}$O obtained using different nucleon potentials (AV4', AV6', Minnesota)
	and different hyperon-nucleon interaction (two-body alone and two-body plus three-body). 
	In the last line the experimental $B_\Lambda$ for $^5_\Lambda$He is from Ref.~\cite{Juria:1973}.
	Since no experimental data for $^{17}_{~\Lambda}$O exists, the reference separation energy is the
	semiempirical value reported in Ref.~\cite{Usmani:1995}.}
	\label{tab:He5L-O17L}
	\end{center}
\end{table}

\begin{figure}[ht]
	\begin{center}
		\includegraphics[width=\linewidth]{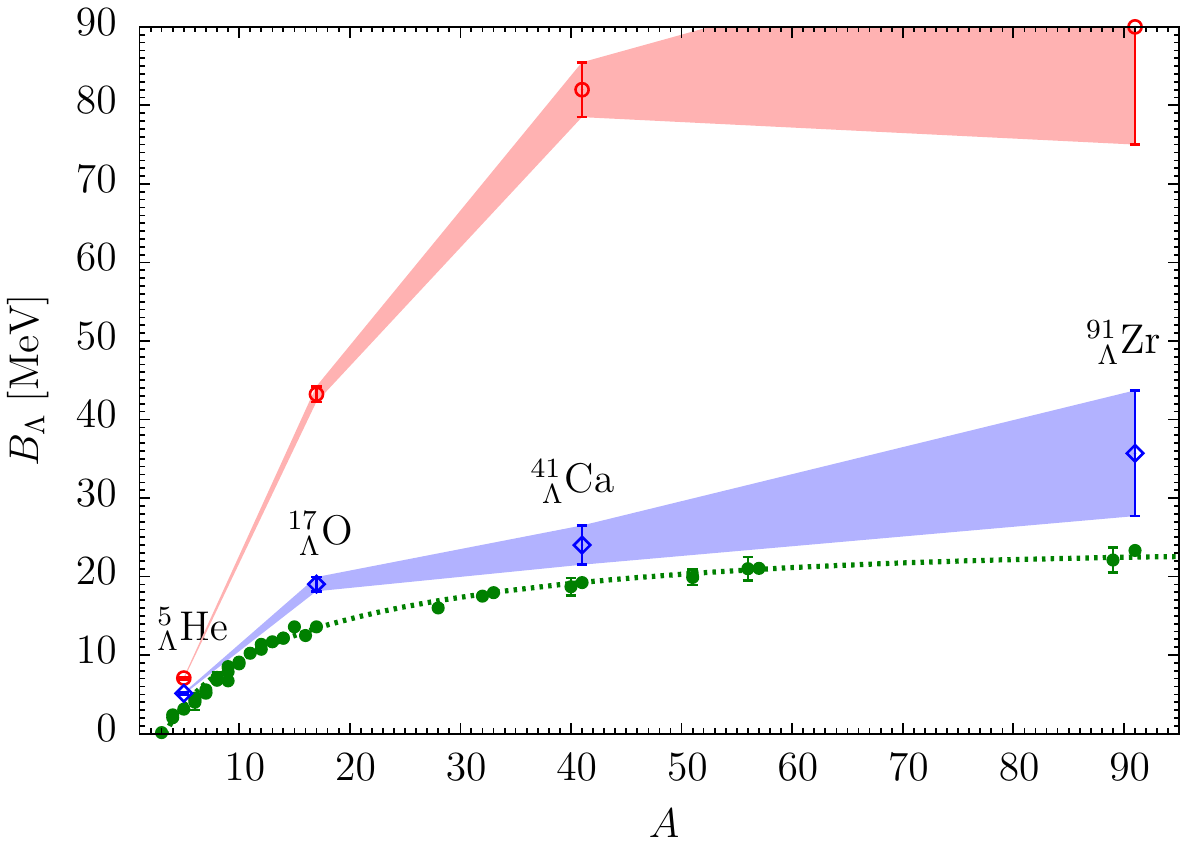}
		\caption{$\Lambda$-separation energy as a function of the baryon number $A$. 
			Plain green dots~(dashed curve) are the available $B_\Lambda$ experimental values. 
			Empty red dots~(upper banded curve) refer to the AFDMC results for the nuclear AV4' 
			potential plus the two-body $\Lambda N$ interaction alone. 
			Empty blue diamonds~(lower banded curve) are the results with the inclusion of the
			three-body hyperon-nucleon force.}
		\label{fig:BL-A_cs}	
	\end{center}
\end{figure}

The results on the $\Lambda$-separation energies are summarized
in Fig.~\ref{fig:BL-A_cs}. We compare the prediction of the hyperon
binding energy in the AV4'+$\Lambda N$ and AV4'+$\Lambda N$+$\Lambda NN$ models for a few 
closed-shell hypernuclei with the experimental values observed in the same mass
range. While the results for lighter hypernuclei might be inconclusive
in terms of the physical consistency of the $\Lambda NN$ contribution to the
hyperon binding energy, the computations for $^{41}_{~\Lambda}$Ca and
$^{91}_{~\Lambda}$Zr reveal a completely different picture. The saturation
binding energy provided by the $\Lambda N$ force alone is completely unrealistic,
while the inclusion of the $\Lambda NN$ force gives results that are qualitatively
much closer to the experimental behavior. We should notice that the
results might be further improved by a refitting of the terms in the
$\Lambda NN$ force. In particular, according to Ref.~\cite{Usmani:1995}, in the
present calculations the $s$-wave contribution is not present. Moreover,
we are missing the explicit inclusion of the kaon exchange term. This
contribution [see Eq.~(\ref{eq:V_YN})] can be estimated at first order
in perturbation theory by computing the expectation of the corresponding
term. As an example, the values of the correction on the total energy
we obtained for $\varepsilon=0.1$~\cite{Usmani:2008}  is -0.33(6)~MeV
in $^5_\Lambda$He and +0.2(4)~MeV in $^{17}_{~\Lambda}$O, the latter
negligible compared to the corresponding binding energy. 

For $^{91}_{~\Lambda}$Zr we should also consider a charge symmetry breaking (CSB) potential. The latter can be easily included as a term of the form
\begin{equation}
	v_{\Lambda i}^{\scriptscriptstyle{\text{CSB}}}=\tau_i^z\,C_0^{\scriptscriptstyle{\text{CSB}}}\,T_\pi^2\left(m_\pi r_{\Lambda i}\right)\,,
	\label{eq:CSB}
\end{equation}
amounting to an isospin-dependent correction to the central potential. 
The inclusion of the CSB term using perturbation theory would be 
zero in isospin-symmetric hypernuclei.
The value $C_0^{\scriptscriptstyle{\text{CSB}}}=-0.050(5)$~MeV reported
in literature~\cite{Usmani:1999}, is fitted in order to reproduce the
difference in $\Lambda$-separation energy of the $A=4$ mirror hypernuclei
($^4_\Lambda$H and $^4_\Lambda$He). According to Eq.~(\ref{eq:CSB}),
the contribution of the charge symmetry breaking term depends on the
difference between the number of neutrons and protons. For $N>Z$ the CSB term is strictly
positive. This implies a repulsive contribution per neutron excess that
would further lower the $B_\Lambda$ for $^{91}_{~\Lambda}$Zr, where there
are 10 more neutrons than protons, thereby reducing the discrepancy with
the experimental result.

In this paper we have presented the first accurate calculation of the 
$\Lambda$-separation energy for closed-shell
$\Lambda$ hypernuclei using the available microscopic interactions. Using
the AFDMC method we were able to extend the calculation in the medium-heavy
range of hypernuclei up to $A=91$, providing, for the first time,
a consistent calculation of light and heavy hypernuclei.
The main outcome of the study is that the inclusion of the three-body $\Lambda NN$
interaction is fundamental in order to reproduce the saturation properties
of the $\Lambda$ binding energy in hypernuclei. The leading contribution to the
three-body interaction is strictly repulsive in the range of hypernuclei
studied. Within the model that we have studied, the inclusion of
the $\Lambda N$ force without a three-body force gives a very 
unphysical $\Lambda$ binding energy.

We speculate that this would lead to a stiffer EOS for the $\Lambda$-neutron
matter when the presented interaction is applied to the study of the
homogeneous medium. This fact might eventually reconcile the onset of
hyperons in the inner core of a neutron star with the observed masses of order
$2M_\odot$. A study along this direction is in progress.

\emph{Acknowledgments.}
Initial stages of this work were performed by 
P. Armani as part of his Ph.D. thesis.  
We thank P.~Armani, S.~Reddy, B.~F.~Gibson, and J.~Carlson for valuable discussions.  
This work was performed partly at LISC, Interdisciplinary Laboratory for
Computational Science, a joint venture of the University of Trento and
Bruno Kessler Foundation. 
Support and computer time were made available by the AuroraScience project (funded by the
Autonomous Province of Trento and INFN) and by Los Alamos Open Supercomputing.
This research also used resources of the National Energy Research
Scientific Computing Center (NERSC), which is supported by the Office of
Science of the U.S. Department of Energy under Contract No. 
DE-AC02-05CH11231.
The work of S.G. is supported by the Nuclear Physics program at the
DOE Office of Science, UNEDF and NUCLEI SciDAC programs and by the LANL
LDRD program.

\bibliographystyle{apsrev4-1}

%

\end{document}